\documentclass[structabstract]{aa}  
\usepackage{graphicx}
\usepackage{txfonts}
\usepackage{natbib}
\begin{document}
   \title{C, N, O abundances and carbon isotope ratios in evolved stars of the open clusters Collinder\,261 and 
   NGC\,6253\thanks{Based on observations collected at ESO telescopes under 
programmes 65.N-0286, 169.D-0473
}
}

   \author{\v{S}ar\=unas Mikolaitis\inst{1},
          Gra\v{z}ina Tautvai\v{s}ien\.{e}\inst{1},
          Raffaele Gratton\inst{2},
          Angela Bragaglia\inst{3}, 
          \and
          Eugenio Carretta\inst{3}
          }
   \institute{Institute of Theoretical Physics and Astronomy, Vilnius University,
              A. Gostauto 12, LT-01108 Vilnius\\
              \email{sarunas.mikolaitis@tfai.vu.lt, grazina.tautvaisiene@tfai.vu.lt}
         \and
        INAF - Osservatorio Astronomico di Padova, Vicolo dell'Osservatorio 5, I-35122 Padova, Italy\\
        \email{raffaele.gratton@oapd.inaf.it}
        \and
        INAF - Osservatorio Astronomico di Bologna, Via Ranzani 1, I-40127 Bologna, Italy\\
        \email{angela.bragaglia@oabo.inaf.it, eugenio.carretta@oabo.inaf.it}
             }
   \date{Received January 17, 2012; accepted March 19, 2012}

\authorrunning{\v{S}. Mikolaitis, G. Tautvai\v{s}ien\.{e}, R. Gratton, A. Bragaglia, and E. Carretta}
\titlerunning{Collinder\,261 and NGC\,6253}

 
  \abstract
   {Investigations of abundances of carbon and nitrogen in the atmospheres of evolved stars 
of open clusters may provide comprehensive information on chemical composition changes caused by 
stellar evolution. }
   {Our main aim is to increase the number of open clusters with determined carbon-to nitrogen and  carbon isotope ratios. }
   {High-resolution spectra were analysed using a differential model atmosphere method. 
Abundances of carbon were  derived using the ${\rm C}_2$ Swan (0,1) band head at
5635.5~{\AA} (FEROS spectra) and the ${\rm C}_2$  Swan (1,0) band head at
4737~{\AA} (UVES spectra). The wavelength interval 7980--8130~{\AA}, with strong
CN features was analysed to determine  nitrogen abundances and
$^{12}{\rm C}/^{13}{\rm C}$  isotope ratios.  The oxygen abundances were
determined from the [O\,{\sc i}] line at 6300~{\AA}.  
}
   {The average value of $^{12}{\rm C}/^{13}{\rm C}$\  isotope ratios of Cr\,261 is equal to $18\pm 2$\ in four giants
 and to $12\pm 1$ in two clump stars; it is equal to $16\pm 1$ in four clump stars of the open cluster NGC\,6253. 
   The mean C/N ratios in Cr\,261 and 
   NGC\,6253 are equal to $1.67\pm 0.06$ and  $1.37\pm 0.09$, respectively. }
  {The $^{12}{\rm C}/^{13}{\rm C}$ and C/N values in Cr\,261 and NGC\,6253 within limits of uncertainties agree 
  with the theoretical model of thermohaline-induced mixing as well as with the  cool-bottom processing model. }

   \keywords{stars: abundances --
                stars: horizontal branch --
                stars: evolution --
                open clusters and associations: individual: Cr~261, NGC~6253
               }

   \maketitle

\section{Introduction}

The carbon and nitrogen abundances, C/N, and especially the carbon isotope 
ratios $^{12}{\rm C}/^{13}{\rm C}$ are key tools for stellar evolution studies.  
It is well known that low-mass stars experience the first dredge-up at the bottom of the giant branch (\citealt{Iben1965}). 
However, extra-mixing processes become efficient on the red giant branch (RGB) when 
these stars reach the so-called RGB  bump, and modify the surface 
abundances (\citealt{Gilroy1989, Gilroy1991, Luck1994, Charbonnel1994, Charbonnel1998, 
Gratton2000, Tautvaisiene2000, Tautvaisiene2001, Tautvaisiene2005, Smiljanic2009, Mikolaitis2010, Mikolaitis2011A, 
 Mikolaitis2011B}). It is also known that alterations of $^{12}{\rm C}/^{13}{\rm C}$\ and 
$^{12}{\rm C}/^{14}{\rm N}$\ ratios depend on stellar evolutionary stage, mass and 
metallicity (\citealt{Charbonnel1998, Gratton2000, Chaname2005, 
Charbonnel2006, Cantiello2010, Charbonnel2010}).  However, details of these dependences
are uncertain; a comprehensive and statistically significant observational 
data base for stars of different turn-off masses and metallicities is needed to constrain models. 
In this work, our targets of investigations are the open clusters Collinder\,261 and NGC\,6253. 
    
Collinder\,261 (Cr\,261; galactic coordinates $l = 301^{\circ}.68, 
b = -5^{\circ}.53$) is one of the oldest open clusters in the Galaxy. Its age is 5--10~Gyr, depending on the adopted 
stellar evolution models (\citealt{Janes1994, Mazur1995MNRAS, Gozzoli1996, Carraro1998}), but the recent investigation of \cite{Bragaglia2006} 
has defined a most probable age of about 6 Gyr. Unlike the majority of 
old open clusters, which are located in the outer Galactic disk, this cluster is in the inner part, 
at 7.5~kpc from the Galactic centre and 235~pc below the plane. Despite the old age, the metallicity of this 
cluster determined from high-resolution spectral studies is close to solar:  ${\rm [Fe/H]}=-0.03$ (\citealt{Carretta2005, 
DeSilva2007}), $-0.22$~dex (\citealt{Friel2003}), $+0.13$~dex (\citealt{Sestito2008}). \citeauthor{DeSilva2007}, who investigated 
12 red giants in Cr\, 261 using Very Large Telescope (VLT) UVES spectra, emphasised that this cluster is very chemically 
homogeneous -- the intrinsic scatter was estimated to be less than 0.05~dex. The chemical homogeneity of Cr\, 261 makes 
this cluster an extremely interesting target for investigating the mixing of sensitive elements such as carbon isotopes and 
nitrogen, which have not been studied in previous studies. 
 
NGC\,6253 (galactic coordinates $l = 335^{\circ}.45, b = -6^{\circ}.26$) is also quite old of 3--5~Gyr 
(\citealt{Piatti1998, Bragaglia2006, Montalto2009}) and located in the inner part of the Galaxy, at 6.6~kpc 
from the Galactic centre. The most interesting signature of 
this open cluster is its high metallicity. Recent high-resolution spectral determinations of [Fe/H] are the 
following: $+0.43$~dex (\citealt{Anthony-Twarog2010}), $+0.36$~dex (\citealt{Sestito2007}), 
$+0.46$~dex (\citealt{Carretta2007}). Since mixing processes 
depend on stellar metallicities, it is important to have C/N and $^{12}{\rm C}/^{13}{\rm C}$ ratios determined in metal-rich  
evolved stars. By now, less than a handful of open clusters with similarly 
high metal-abundance have been detected in our Galaxy (e.g., NGC~6791, \citealt{gratton06}; NGC~2632, \citealt{pace08}).
This makes NGC\,6253 a very attractive target for our study. 
 
Our main aim is to determine detailed elemental abundances of carbon,
nitrogen and oxygen, and carbon isotopic $^{12}{\rm C}/^{13}{\rm C}$ ratios  in evolved stars of Cr\,261 and NGC\,6253 
to better understand abundance alterations caused by stellar evolution.

  
\input epsf
\begin{figure}
\epsfxsize=\hsize 
\epsfbox[-5 -10 930 850]{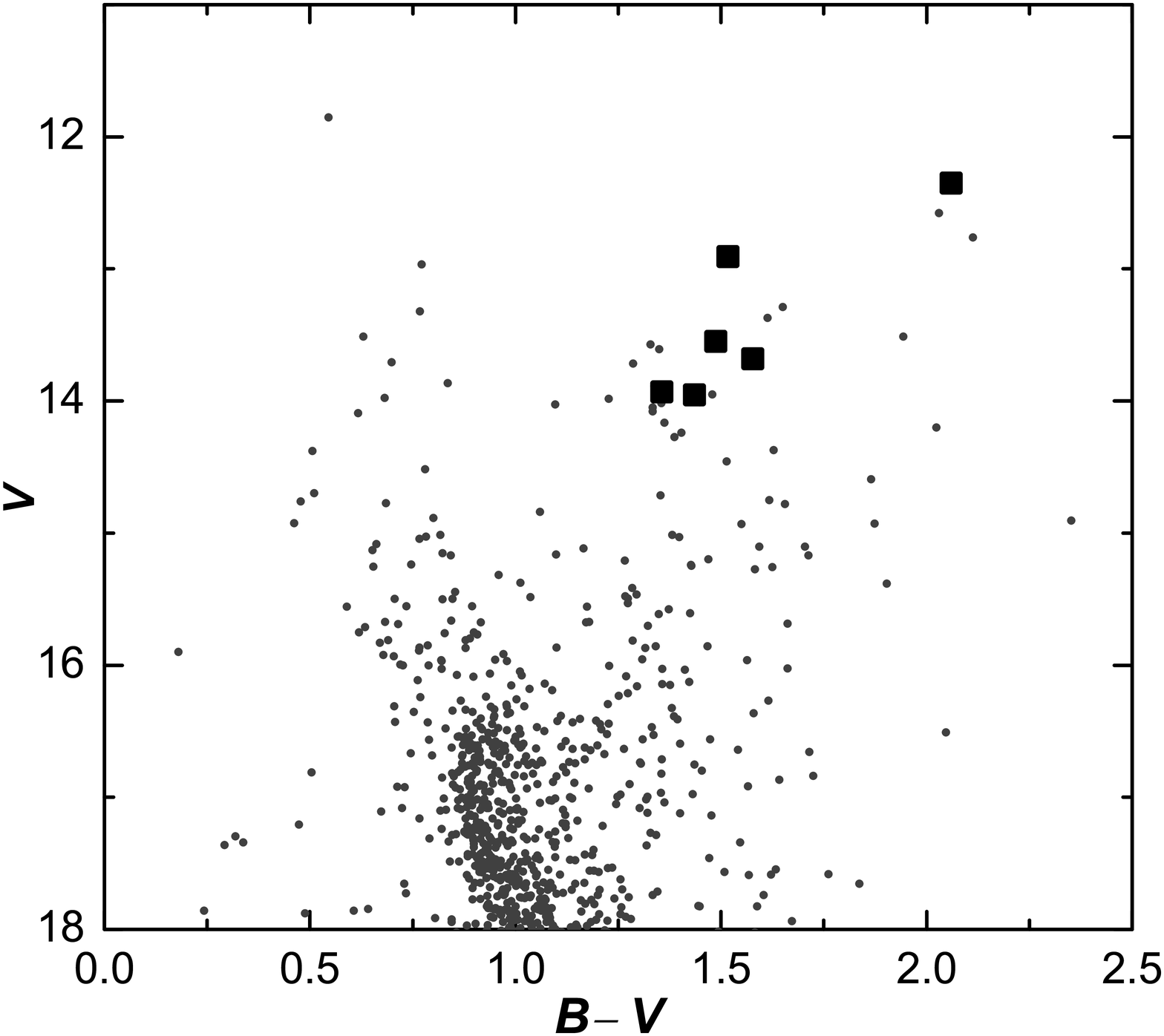} 
\caption{Colour-magnitude diagram of the open cluster Cr\,261. The stars investigated in this work are 
indicated by the filled squares. 
The diagram is based on $BVI$\ photometry by \citet{Gozzoli1996}.} 
\label{Fig1}
\end{figure}
 
  
\input epsf
\begin{figure}
\epsfxsize=\hsize 
\epsfbox[-5 -10 930 850]{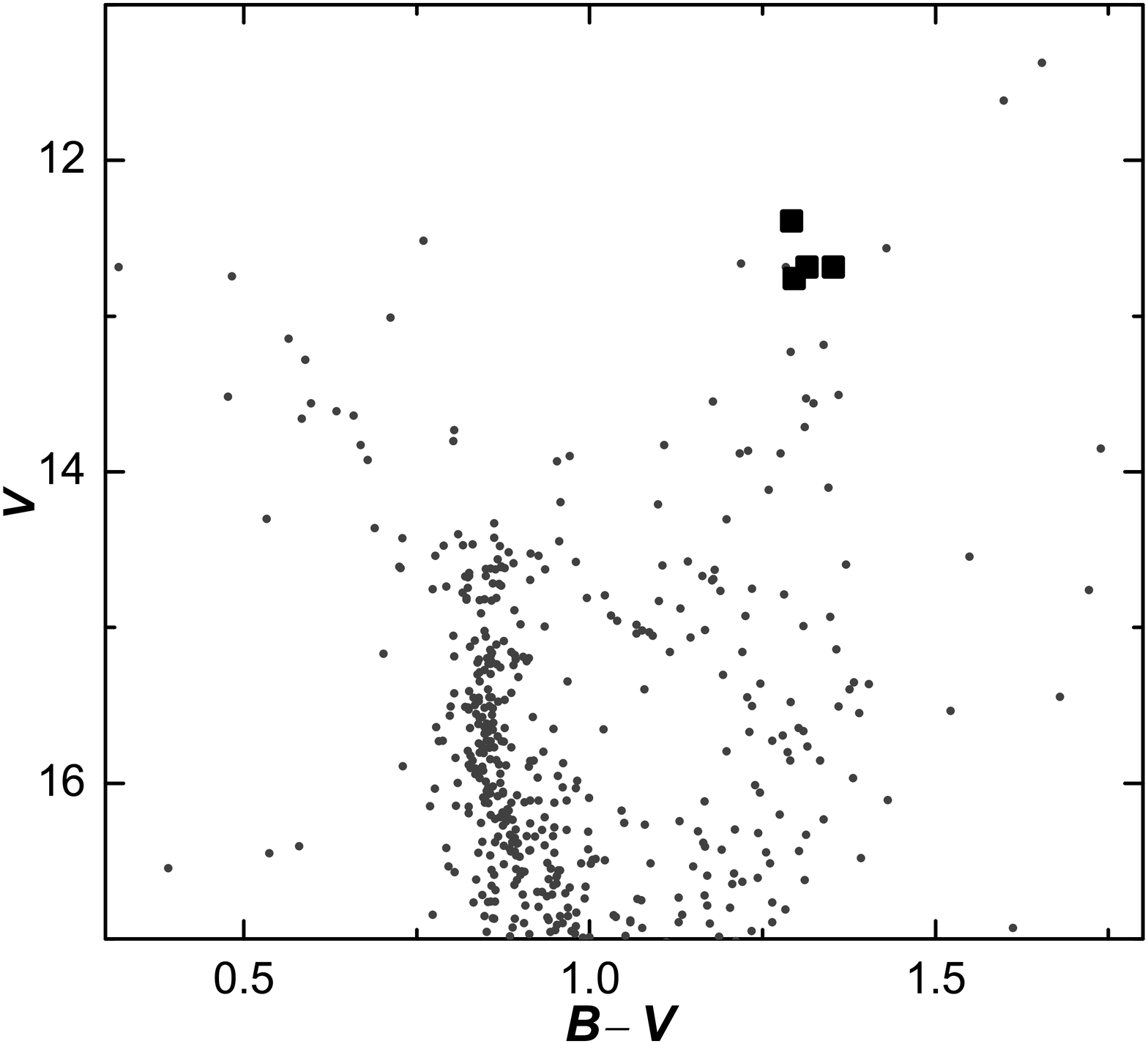} 
\caption{Colour-magnitude diagram of the open cluster NGC\,6253. The stars investigated in this work are 
indicated by filled squares. 
The diagram is based on  $UBVI$\ photometry by  \citet{Bragaglia1997}.} 
\label{Fig2}
\end{figure}

\section{Observations and method of analysis}

The spectra of six cluster stars of Cr\,261 (1045, 1080, 1485, 1871, 2001, and 2105; the identification numbers 
are from \citealt{Phelps1994}) were obtained with the spectrograph FEROS (Fiber-fed Extended Range Optical 
Spectrograph) mounted on the 1.5~m telescope at the European Southern Observatory (ESO) in La Silla (Chile). 
The resolving power is $R=48\,000$ and the wavelength range is 3700$<\lambda<$8600\,{\AA}. 
Two stars (1080 and 2001) belong to the red clump of the cluster, three stars (1045, 1485 and 2105)   
are first-ascent giants, and the star 1871 is an RGB-tip giant (see Fig.~\ref{Fig1}). Depending on the magnitudes, 
signal-to-noise ratios of the observed spectra lie between 130 and 70. The log of observations and 
information about reductions  were presented in the paper by \citet{Carretta2005}. 

The spectrum of NGC\,6253~3595 was obtained with the same spectrograph FEROS.
In the same cluster, the stars 2509, 2885, and 4510 were observed using the UVES spectrograph on the ESO VLT 
telescope in Paranal (Chile). The identification numbers 
are from \citet{Bragaglia1997}. The spectra cover the wavelength ranges 3560--4840 and 5710--9320~\AA\ at 
$R=43\,000$.  Signal-to-noise ratios of the UVES spectra lie between 180 and 120. The log of observations and 
information about reductions  were presented in the paper by \citet{Carretta2007}. \citeauthor{Carretta2007} have 
also observed the star NGC\,6253~2508, but its radial velocity was  
different from the rest of the stars by about $10\sigma$. Even though \citet{Montalto2009} give a high membership probability for this star 
based on its proper motion, we neglected NGC\,6253~2508 in the analysis.
All stars observed in NGC\,6253 belong to the red clump of the cluster (see Fig.~\ref{Fig2}). 

In this work as well as in other papers of this series (\citealt{Mikolaitis2010, Mikolaitis2011A, Mikolaitis2011B}), the  {\sc atlas} models 
with overshooting (\citealt{Kurucz1993}) and a computing code by  \citet{Gratton1988} were used for the analysis 
of the spectra. The analysis was differential to the Sun.

The main atmospheric parameters for the observed stars in Cr\,261 
have been determined spectroscopically  by \citet{Carretta2005}. We used  them in our analysis and  present them in 
Table~\ref{table:1} for convenience.

For the cluster NGC\,6253 the atmospheric parameters have been previously determined by \citet{Carretta2007}, 
but they used a photometric method. To make 
our study as much homogeneous as possible, we decided to redetermine them spectroscopically.

The spectroscopic effective temperatures were derived by minimising the slope of the abundances obtained from 
neutral Fe\,{\sc i} lines with respect to the excitation potential. 
The gravities (log~$g$) were derived by forcing measured neutral and ionised iron lines to 
yield the same [Fe/H] value by adjusting the model gravity.  
The microturbulent velocities were determined by forcing Fe\,{\sc i} abundances to be independent of the equivalent 
widths of lines. 
Iron lines were restricted to the spectral range 5500--8000\,{\AA} to minimise 
problems of line crowding and difficulties in the continuum tracing in the blue region. After the careful selection, 
the number of Fe\,{\sc i} lines was set to 63 and of Fe\,{\sc ii} to 5. First, using the model atmosphere of the Sun 
from the same grid of \citet{Kurucz1993} and the microturbulent velocity of 0.9~km\,s$^{-1}$, we calculated 
the solar iron abundance. The atomic parameters and solar equivalent widths of lines were taken from 
\citet{Gurtovenko1989}. These solar iron abundances were used for the differential determination of abundances 
in the programme stars.
  
The determined atmospheric parameters and iron abundances for the observed stars 
in NGC\,6253 are presented in Table~1. Differences between the photometric and spectroscopic parameters appeared 
to be quite small and lie within uncertainties of determinations.  There is no systematic difference in $T_{\rm eff}$, just 
a mean scatter of $\pm30$~K, our log~$g$ values are by about 0.07~dex higher, $v_{\rm t}$ by about 0.2~km\,s$^{-1}$ higher,  
and differences in [Fe/H] do not exceed $\pm 0.05$~dex.

In this work, abundances  of $^{12}{\rm C}$, $^{13}{\rm C}$, N, and O were determined using the same method 
of analysis as in \citeauthor{Mikolaitis2010} (\citeyear{Mikolaitis2010}, Paper~I). Here we recall only some details. 

For the carbon abundance determination in stars observed with FEROS, we considered observations of the 
5632 -- 5636~{\AA} spectral range close to the ${\rm C}_2$ Swan (0,1) band head.
This spectral interval was not available for the stars observed with UVES, therefore we analysed several lines of the 
Swan (1,0) bands at 4732.8~{\AA} and 4735.3~{\AA}. This spectral interval was available in the FEROS spectra 
as well, but we were unable to analyse the ${\rm C}_2$ lines at these wavelengths since the S/N ratio was too low.

The interval 7980 -- 8130~{\AA} containing strong $^{12}{\rm C}^{14}{\rm N}$ and $^{13}{\rm C}^{14}{\rm N}$ 
features was used for the nitrogen abundance and $^{12}{\rm C}/^{13}{\rm C}$ ratio analysis. 

We derived the oxygen abundance from synthesis of the forbidden [O\,{\sc i}] line at 6300~{\AA}. 
The $gf$\ values for $^{58}{\rm Ni}$ and $^{60}{\rm Ni}$ isotopic line components, which blend the 
oxygen line, were taken from \citet{Johansson2003}. 
The [O\,{\sc i}] line was not contaminated 
by telluric lines in the spectra of investigated stars. 
The solar abundances were taken from \citet{Grevesse2000}. 
The carbon, nitrogen and oxygen abundances used in our work are $\log{A_{\rm C}} = 8.52$, 
$\log{A_{\rm N}} = 7.92$, and $\log{A_{\rm O}} = 8.83$. The solar C/N ratio is 3.98. 
The $^{12}{\rm C}/^{13}{\rm C}$\ ratio in the solar photosphere is equal to 89 
(\citealt{Coplen2002}).  All synthetic spectra were first calibrated to the solar spectrum by \citet{Kurucz2005}.

Figures 3, 4, 5, and 6 display examples of spectrum syntheses for the programme stars. The best-fit abundances were 
determined by eye.

\input epsf
\begin{figure}
\epsfxsize=\hsize 
\epsfbox[-20 -20 830 650]{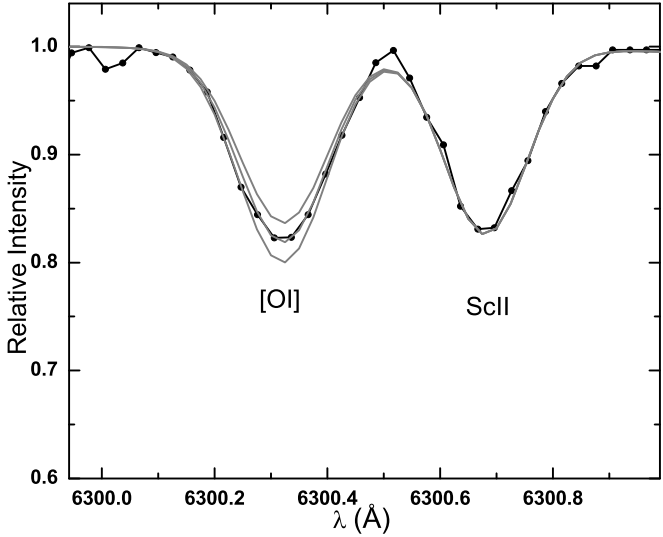} 
    \caption {Fit to the forbidden [O\,{\sc i}] line at 6300~{\AA} in 
Cr\,261\,1080. The observed spectrum is shown as a solid line with black dots. Synthetic
spectra with [O/Fe]$=0.05$, $-0.05$, and $-0.15$ are shown as solid grey lines.} 
    \label{Fig3}
  \end{figure}

\input epsf
\begin{figure}
\epsfxsize=\hsize 
\epsfbox[-20 -20 830 670]{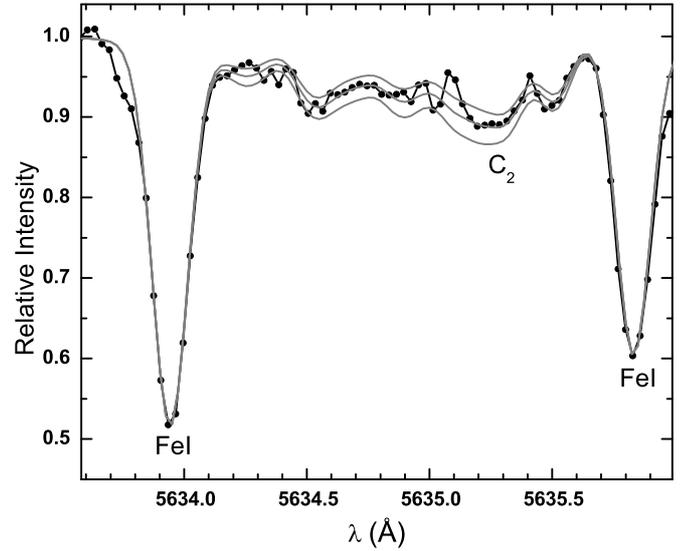} 
    \caption {Small region of Cr\,261 2001 spectrum (solid black line with black dots) at 
${\rm C}_2$ Swan (0,1) band head 5635.5~{\AA}, plotted together with 
synthetic spectra with [C/Fe] values of $-0.10$~dex (lower grey line), $-0.15$~dex (middle grey line) 
and $-0.20$ (upper grey line). 
}
    \label{Fig4}
  \end{figure}

\input epsf
\begin{figure}
\epsfxsize=\hsize 
\epsfbox[-20 -20 830 670]{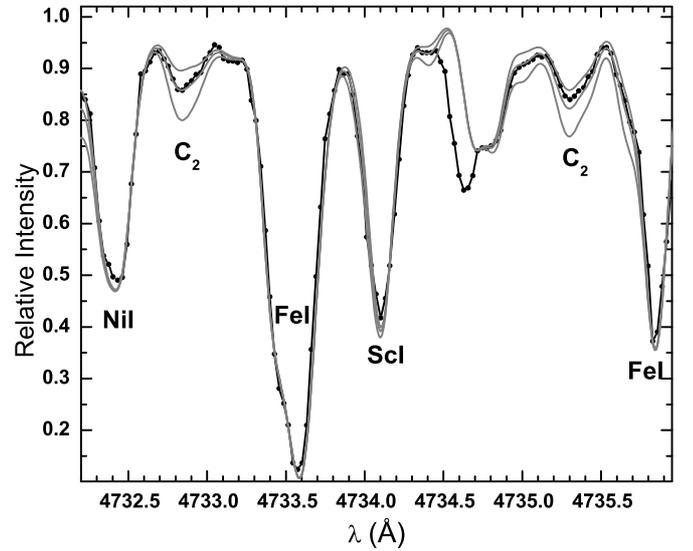} 
    \caption {Small region of NGC\,6253\,2885 spectrum (solid black line with black dots) with 
${\rm C}_2$ features, plotted together with 
synthetic spectra with [C/Fe] values of $-0.05$~dex (lower grey line), $-0.15$~dex (middle grey line) 
and $-0.25$ (upper grey line). 
}
    \label{Fig5}
  \end{figure}

  \begin{table*}
     \centering
\begin{minipage}{190mm}
\caption{Main parameters and elemental abundances of the programme stars.}
\label{table:1}
\begin{tabular}{llcccccrrrrccrcc}
\hline\hline
Cluster\tablefootmark{1, 2} & Star & $V$ & $B-V$ & $T_{\rm eff}$ & 
log~$g$ & $v_{\rm t}$ & [Fe/H] & [C/Fe]& [N/Fe] & $\sigma$\tablefootmark{3}& 
n\tablefootmark{4}& [O/Fe]& C/N& $^{12}{\rm C}/^{13}{\rm C}$\\
& &  mag & mag & K & & km s$^{-1}$ & & & & & & & &     \\
  \hline
\noalign{\smallskip}
   Cr~261 &    1045 &    13.547 &    1.487 &    4470 &    2.07 &    1.23 
&    0.00 &    --0.15 &    0.23 &    0.05 &    16 &    --0.05 &    1.66 
&    15    \\
   Cr~261 &    1080 &    13.952 &    1.435 &    4500 &    2.09 &    1.23 
&    0.00 &    --0.15 &    0.26 &    0.05 &    10 &    --0.05 &    1.58 
&    11    \\
   Cr~261 &    1485 &    13.680 &    1.577 &    4340 &    1.76 &    1.27 
&    --0.06 &    --0.15 &    0.23 &    0.05 &    16 &    --0.05 &    
1.66 &    18    \\
   Cr~261 &    1871 &    12.350 &    2.060 &    3980 &    0.43 &    1.44 
&    --0.31 &    --0.10 &    0.20 &    0.04 &    20 &    --0.20 &    
2.04 &    18    \\
   Cr~261 &    2001 &    13.932 &    1.356 &    4580 &    1.83 &    1.26 
&    --0.02 &    --0.15 &    0.20 &    0.07 &    10 &    --0.10 &    
1.66 &    13    \\
   Cr~261 &    2105 &    12.908 &    1.517 &    4180 &    1.59 &    1.29 
&    --0.08 &    --0.10 &    0.24 &    0.06 &    16 &    --0.05 &    
1.78 &    19    \\
\noalign{\smallskip}
\hline
\noalign{\smallskip}
NGC~6253 &    2509 &    12.685 &    1.314 &    4494 &    2.57 &    1.42 
&    0.46 &    --0.17 &    0.35 &    0.03 &    21 &    --0.10 &    1.20 
&    15    \\
NGC~6253 &    2885 &    12.656 &    1.352 &    4490 &    2.43 &    1.38 
&    0.43 &    --0.23 &    0.33 &    0.05 &    21 &    --0.20 &    1.12 
&    17    \\
NGC~6253 &    3595 &    12.388 &    1.292 &    4535 &    2.44 &    1.40 
&    0.44 &    --0.18 &    0.27 &    0.04 &    19 &    --0.10 &    1.39 
&    17    \\
NGC~6253 &    4510 &    12.759 &    1.296 &    4509 &    2.52 &    1.38 
&    0.47 &    --0.20 &    0.32 &    0.03 &    21 &    --0.17 &    1.20 
&    15    \\
\hline
\end{tabular}
\end{minipage}
\tablefoot{
\tablefoottext{1}{Star numbers  for the cluster Cr\,261 were taken from 
\citet{Phelps1994} and $V$, and $B-V$ are from \citet{Gozzoli1996}.
Star numbers, $V$, $B-V$ for the cluster NGC~6253 were used from 
\citet{Bragaglia1997}}
\tablefoottext{2}{The main atmospheric parameters for the cluster
Cr~261 were adopted from \citet{Carretta2005}, for the cluster NGC~6253 
-- determined in this work. }
\tablefoottext{3}{The standard deviations in the mean value due to the 
line-to-line scatter.}
\tablefoottext{4}{The number of CN molecular features used.}}
\end{table*}

\input epsf
\begin{figure}
\epsfxsize=\hsize 
\epsfbox[-20 -20 830 700]{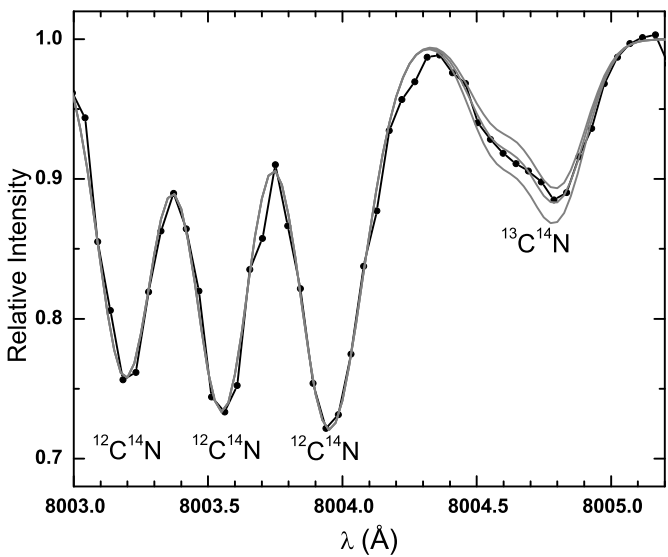} 
    \caption {Small region of NGC\,6253\,4510 spectrum (solid black line with black dots) 
    including a $^{13}{\rm C}^{14}{\rm N}$ feature.
Grey lines show synthetic spectra with $^{12}{\rm C}$/$^{13}{\rm C}$ ratios equal to 12 (lower line), 
15 (middle line) and 18 (upper line). 
}
    \label{Fig6}
  \end{figure}

\subsection{Estimation of uncertainties}

Sources of uncertainty in similar analyses were described in detail and evaluated in our Paper~I;
that discussion applies also to the stars and instrumentation used here for Cr\,261. 

Since stars in the open cluster NGC\,6253 are more metal-abundant than those in other cluster analysed by us 
previously, it was interesting to see what the sensitivities to uncertainties of the main atmospheric 
parameters are.  The sensitivity of the abundance 
estimates to variations in the atmospheric parameters for the assumed parameter differences of
$\pm~100$~K for $T_{\rm eff}$, $\pm 0.3$~dex for $\log{g}$,  
$\pm 0.3~{\rm km~s}^{-1}$ for $v_{\rm t}$\ , and $\pm 0.1$~dex for [Fe/H] is 
illustrated in Table~\ref{table:2} for the star NGC\,6253\,2509. Evidently, possible 
parameter errors do not affect the abundances seriously; the element-to-iron 
ratios, which we use in our discussion, are even less affected. 
For Cr\,261, the sensitivity of iron abundances to stellar atmospheric parameters were described 
in \citet{Carretta2005}.

Since abundances of C, N, and O are bound together by the molecular equilibrium 
in the stellar atmosphere, we also investigated how an error in one of 
them typically affects the abundance determination of another. 
$\Delta{\rm [O/H]}=0.10$ causes 
$\Delta{\rm [C/H]}=0.05$ and $\Delta{\rm [N/H]}=-0.15$;   
$\Delta{\rm [C/H]}=0.10$ causes $\Delta{\rm [N/H]}=-0.15$ and 
$\Delta{\rm [O/H]}=0.05$; 
$\Delta {\rm [N/H]}=0.10$ has no effect on either the carbon or the oxygen abundances.

The random errors of elemental abundance determinations mostly depend on the S/N ratio and can be evaluated from 
the star-to-star scatter and line-to-line scatter for an individual star. The star-to-star scatter is very small for the 
investigated cluster stars, it does not exceed  0.07~dex. The line-to-line scatter in our work  is seen in the case of nitrogen 
 abundance determination (Table~\ref{table:1}); its mean value is equal to $0.05$~dex for Cr\,261, and to $0.04$~dex for 
 NGC\,6253.  The typical errors of $^{12}{\rm C}/^{13}{\rm C}$ and C/N ratios are 2.5 and 0.25, respectively 
 (cf. \citealt{Charbonnel2010, Smiljanic2009,  Gilroy1989}).

    \begin{table}
    \centering
       \caption{Effects on derived abundances, $\Delta$[A/H], resulting from model changes
for the star NGC\,6253\,2509. }
         \label{table:2}
       \[
          \begin{tabular}{lrrccc}
             \hline
             \noalign{\smallskip}
Species & ${ \Delta T_{\rm eff} }\atop{ \pm100~{\rm~K} }$ &
             ${ \Delta \log g }\atop{ \pm0.3 }$ &
             ${ \Delta v_{\rm t} }\atop{ \pm0.3~{\rm km~s}^{-1}}$ &
             ${ \Delta {\rm [Fe/H]} }\atop{ \pm0.1}$ &
             Total \\
             \noalign{\smallskip}
             \hline
             \noalign{\smallskip}
C\,(C$_2$) &    0.07 &    0.05 &    0.00 &    0.03 &    0.09    \\
N\,(CN) &    0.07 &    0.05 &    0.03 &    0.05 &    0.10    \\
O\,([O\,{\sc i}]) &    0.03 &    0.03 &    0.00 &    0.03 &    0.05    \\
$^{12}$C/$^{13}$C&    1.5 &    1.5 &    0.0 &    0.00 &    2    \\
                  \noalign{\smallskip}
             \hline
          \end{tabular}
       \]
    \end{table}

\section{Results and discussion}

We determined the abundances of the key chemical elements 
$^{12}{\rm C}$, $^{13}{\rm C}$, N, and O for the open clusters Cr\,261 and NGC\,6253. 
The abundances relative to iron
[El/Fe]\footnote{In this paper we use the customary spectroscopic notation
[X/Y]$\equiv \log_{10}(N_{\rm X}/N_{\rm Y})_{\rm star} -
\log_{10}(N_{\rm X}/N_{\rm Y})_\odot$} and $\sigma$ (the line-to-line 
scatter) derived for the programme stars are listed in Table~\ref{table:1}.

As already noted by \cite{Carretta2005} (who attributed this to the incomplete adequacy of Kurucz atmospheric 
models for cool giants), the [Fe/H] value of the star Cr\,261\,1871, which is located at the red giant branch-tip, is 
slightly peculiar because it is more than 0.2~dex lower than of other cluster 
stars. Its carbon and nitrogen to iron ratios are indistinguishable 
from other stars of this cluster; but the oxygen abundance and C/N ratio slightly differ 
from the other stars of this cluster. We did not use the [O/Fe] and C/N values of this star when 
we calculated the average values for the cluster.  The value of $^{12}{\rm C}/^{13}{\rm C}$, 
which is less sensitive to the atmosphere parameters, agrees with results for 
other stars. 

The average values of carbon to iron ratios in Cr\,261 and NGC\,6253 stars  are ${\rm [C/Fe]}=-0.13\pm0.02$ 
and  ${\rm [C/Fe]}=-0.20\pm0.01$, respectively. 
This means that, the ratios of [C/Fe] in investigated stars of these clusters lie below 
the values obtained for dwarf stars of the Galactic disk, which are solar. 
\citet{Shi2002} performed an abundance 
analysis of carbon for a sample of 90 F- and G-type main-sequence disk stars  
using C\,{\sc i} and [C\,{\sc i}] lines and found [C/Fe] to be about solar 
at solar metallicity.   
Roughly solar carbon abundances were found by \citet{Gustafsson1999}, 
who analysed a sample of 80 late-F and early-G type dwarfs using the forbidden 
[C\,{\sc i}] line. 

While determining the carbon abundance, we also had to analyse oxygen since carbon and oxygen are 
bound together by the molecular equilibrium in the stellar atmosphere.  
The mean oxygen-to-iron abundance ratios 
in Cr\,261 and NGC\,6253 stars are ${\rm [O/Fe]}=-0.06\pm0.02$ and ${\rm [O/Fe]}=-0.15\pm0.05$, respectively.  
These results agree with the oxygen abundances in metal-rich dwarfs (\citealt{Bensby2004} and references therein). 
The results for C and O also agree well with the values found by \cite{Carretta2005} and \cite{Carretta2007}. 
   
The mean nitrogen-to-iron abundance ratios 
in Cr\,261 and NGC\,6253 stars are ${\rm [N/Fe]}=0.23\pm0.02$ and ${\rm [N/Fe]}=0.29\pm0.03$, respectively.  
This shows that the nitrogen abundances are enhanced in these evolved stars,
since [N/Fe] values in the Galactic main-sequence stars are about solar at the solar
metallicity (c.f. \citealt{Shi2002}). The low [N/Fe] values previously found in  NGC\,6253 by \cite{Carretta2007} were 
caused by an error in input parameters.

Because we compare the abundances of C, N, and O with results for dwarfs from the literature, some offsets might be present owing to 
differences in the analysed spectral features, solar abundance scales, or other parameters. However, because we compare 
element-to-iron ratios, which are less sensitive to systematic errors, our qualitative evaluation of systematic abundance differences 
should be correct.

The final C/N and $^{12}{\rm C}/^{13}{\rm C}$\ ratios achieved by the star after the mixing events depend on  stellar 
turn-off masses, which are $1.10~M_{\odot}$ 
and $1.40~M_{\odot}$ (\citealt{Bragaglia2006}) for  Cr\,261 and NGC\,6253, respectively. 
The mean C/N ratios in Cr\,261 and NGC\,6253 are equal to $1.67\pm 0.06$ and  $1.37\pm 0.09$, respectively. 

For clusters in which both first-ascent giants after the red giant branch (RGB) luminosity bump and 
 clump stars were analysed,  it is worth to check whether the carbon isotope ratios are similar or not. Two investigated stars 
 in Cr\,261 belong to the clump (1080 and 2001) and the remaining stars are giants located above the bump. 
We see that carbon isotope ratios are lowered more in the clump stars than in the giants. 
The mean  $^{12}{\rm C}/^{13}{\rm C}$\ ratio is equal to $12\pm1$ in the clump stars and to $18\pm2$ in the giants. 
The mean $^{12}{\rm C}/^{13}{\rm C}$ ratio in four clump stars investigated in NGC\,6253 is equal to $16\pm1$. 

The C/N is much less sensitive to mixing processes then $^{12}{\rm C}/^{13}{\rm C}$\ ratio 
(\citealt{Boothroyd1999, Charbonnel2010}; and references therein). 
We see this in Fig.~7 and 8.  The C/N ratio in clump stars of Cr\,261 is by 0.1 
lower than in the giants, but this difference lies within the uncertainties of determination.    

It is important to increase the number of open clusters of different age and metal abundance with $^{12}{\rm C}/^{13}{\rm C}$ 
ratios determined in giants and clump stars. This may help us to better understand mixing processes in low-mass stars and 
possible He-flash influence. 
 
\section{Comparison with theoretical models} 
  
The low-mass stars ($M < 2.5 M_{\odot}$) approach the red giant branch after the turn-off from the main sequence.
The convective envelope deepens inside, towards the hydrogen burning shell. The point of the deepest penetration 
is the end of the first dredge-up. A sharp composition discontinuity is left after this event.
The result is the altered surface abundances of $^3{\rm He}$ and of the main mixing tracers Li, Be, B, C, and N,
depending on the initial stellar mass and chemical composition (\citealt{Charbonnel1994, Charbonnel2007, Charbonnel2010, Boothroyd1999}).
After the completion of the first dredge-up, canonical mixing models, where
the rotation is not included and the convection is the only tool of mixing inside the stellar interior,
do not predict any other alterations in the surface
abundances until the stars reach the asymptotic giant branch.
However, because observations of star in the Galactic field and open or globular clusters
showed signatures of extra-mixing process, new mechanisms of extra-mixing were proposed by a 
number of scientific groups to describe the observed surface abundances in various types of stars 
(see reviews by \citealt{Chaname2005, Charbonnel2006}; recent papers by 
\citealt{Denissenkov2010, Lagarde2011, Palmerini2011, Wachlin2011, Angelou2012} and references therein).

\begin{figure}
\epsfxsize=\hsize 
\epsfbox[-20 -20 860 700] {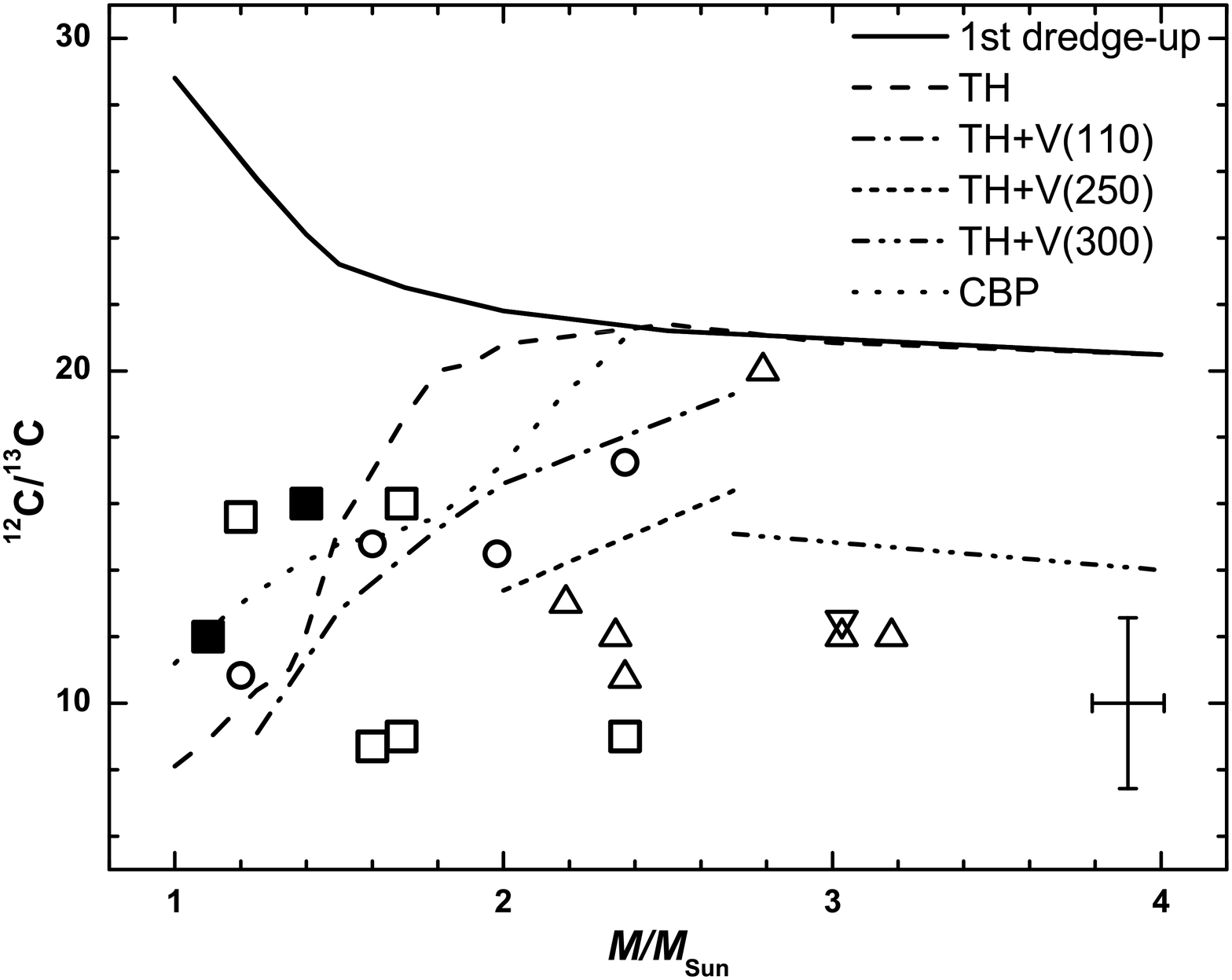} 
    \caption {Average carbon isotope ratios in clump stars of open clusters
as a function of stellar  turn-off mass. 
The results of this work are marked by filled squares; 
from \citet{Mikolaitis2010, Mikolaitis2011A, Mikolaitis2011B} and \citet{Tautvaisiene2000, Tautvaisiene2005} -- open squares; 
from \citet{Smiljanic2009} -- open triangles; from \citet{Luck1994} -- reversed open triangle;
from \citet{Gilroy1989} -- open circles. 
The models of the first dredge-up, thermohaline mixing (TH) and rotation-induced 
mixing (V) are taken from \citet{Charbonnel2010}. The
CBP model of extra-mixing is taken from \citet{Boothroyd1999}. A typical error bar is indicated (\citealt{Charbonnel2010, Smiljanic2009, Gilroy1989}). 
}
    \label{Fig7}
  \end{figure}

\begin{figure}
\epsfxsize=\hsize 
\epsfbox[-20 -20 860 700] {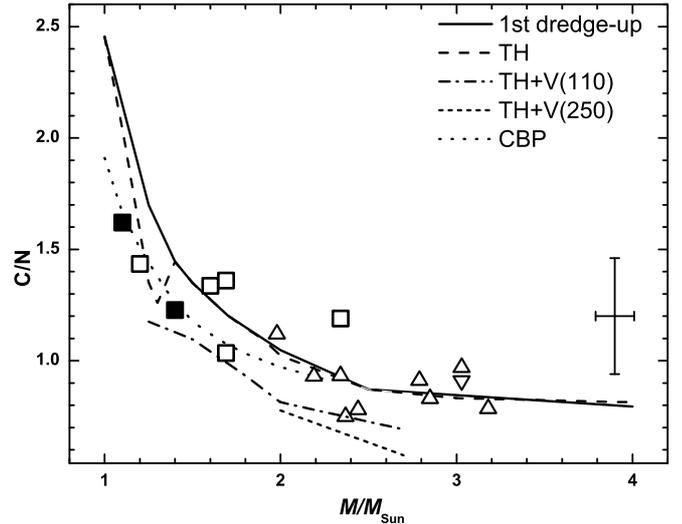} 
    \caption {Average carbon-to-nitrogen ratios in clump stars of open clusters
as a function of stellar  turn-off mass. The meaning of symbols are as in Fig.~\ref{Fig7}.
}
    \label{Fig8}
  \end{figure}

In Figs.~\ref{Fig7} and \ref{Fig8} we compare the determined $^{12}{\rm C}/^{13}{\rm C}$ and C/N ratios in clump 
stars of Cr\,261 and NGC\,6253 
with theoretical models and results obtained for other open clusters.  The mean values of the $^{12}{\rm C}/^{13}{\rm C}$ 
and C/N ratios in the clump stars of Cr\,261 are $12\pm 1$ and $1.62\pm 0.04$, respectively. 
The clump stars have accumulated all chemical composition changes that have happened during 
their evolution along the giant branch and during the helium flash, therefore they are very suitable for a comparison. 
A compilation of recent 
$^{12}{\rm C}/^{13}{\rm C}$\ and C/N ratios in clump stars of open clusters has been provided by \citet{Mikolaitis2011A}. 
In addition, we include the results by \citet{Mikolaitis2011B} and this work. Two models of 
extra-mixing are compared: the thermohaline-induced mixing (TH) model  
(\citealt{Charbonnel2010}), and the cool-bottom processing (CPB) model by \citet{Boothroyd1999}.    

The cool-bottom processing model, which includes a deep circulation mixing 
below the base of the standard convective envelope was proposed more than a decade ago  
(\citealt{Boothroyd1995, Wasserburg1995, Boothroyd1999} and references therein). In this model, 
an extra-mixing takes material from the convective envelope, transports it down to regions hot enough for some nuclear
 processing in the outer wing of the H-burning shell, and then transports it back up to the convective envelope. For the 
 computations of the extra-mixing the "conveyor-belt" circulation model was used. The temperature difference between 
 the bottom of mixing and the bottom of the H-burning shell was considered a free parameter, to be determined by 
 comparison with the observations, to M~67 (\citealt{Gilroy1989, Gilroy1991}), in particular.

The model of thermohaline instability induced mixing is based on ideas of
\citet{Eggleton2006} and \citet{Charbonnel2007}.  \citet{Eggleton2006} found
a mean molecular weight ($\mu$) inversion in their $1~M_{\odot}$ stellar evolution model, occurring after
the so-called luminosity bump on the RGB, when the hydrogen-burning shell reaches the chemically
homogeneous part of the envelope. The $\mu$-inversion is produced by the reaction
$^3{\rm He(}^3{\rm He,2p)}^4{\rm He}$, as predicted by \citet{Ulrich1972}. It does not occur earlier, because the
magnitude of the $\mu$-inversion is low and negligible compared to a stabilising
$\mu$-stratification. Following \citeauthor{Eggleton2006}, \citet{Charbonnel2007} computed stellar models
including the prescription by \citet{Ulrich1972} and extended them to the case of a non-perfect gas for the turbulent
diffusivity produced by that instability in a stellar radiative zone. They found that a double diffusive instability
referred to as thermohaline convection, which has been discussed long ago in the literature (\citealt{Stern1960}),
is important in evolution of red giants. This mixing connects the convective envelope with the external wing of
the hydrogen-burning shell and induces surface abundance modifications in red giant stars.

The thermohaline instability induced mixing  is widely used in oceanology. It is used to model the regions of cooler, less 
salty water below the warmer water where the salinity is higher because of the evaporation from the surface.
The so-called "long fingers" of the warmer water penetrate the cooler water; mixing occurs when the heat excess is 
exchanged (e.g., \citealt{Schmitt1983, Schmitt2003, Ruddick2003, Kunze2003, Radko2010}).

\input epsf

\input epsf

In Figs.~\ref{Fig7} and \ref{Fig8} we can see that the $^{12}{\rm C}/^{13}{\rm C}$ and C/N values 
in Cr\,261 and NGC\,6253 agree within limits of uncertainties 
with the theoretical models of extra-mixing, as well as with other observational studies. 
The mean $^{12}{\rm C}/^{13}{\rm C}$ ratio of stars in NGC\,6253 lie slightly above the 
solar metallicity models in accordance with the theory. However, there are no models computed for supersolar 
metallicities and there are no have other observed clusters with lower metallicities and the same turn-off mass as  NGC\,6253, 
therefore we cannot judge the agreement quantitatively.  In comparison to the most metal-deficient cluster in our sample 
NGC\,2506, which has [Fe/H]$=-0.24$ and turn-off mass 1.7~$M_{\odot}$, the $^{12}{\rm C}/^{13}{\rm C}$ ratio of 
NGC\,6253 is higher by 5.  We need more observational data to determine the sensitivity of extra-mixing on 
metal abundance.  As we already emphasised in our 
previous studies (\citealt{Mikolaitis2011A, Mikolaitis2011B}), the $^{12}{\rm C}/^{13}{\rm C}$ values in the clump stars 
with turn-off masses of 2--3~$M_{\odot}$\ in most of the investigated clusters are lower than predicted by 
the available models and need modelling of stronger extra-mixing. 

\citet{Charbonnel2010} also computed the models of rotation-induced mixing for stars at the zero age main 
sequence (ZAMS) with rotational velocities of 110\,km\,s$^{-1}$, 250\,km\,s$^{-1}$ and 300\,km\,s$^{-1}$. 
Typical initial ZAMS rotation
velocities were chosen depending on the stellar mass based on observed rotation distributions in young open clusters, see
\citet{Gaige1993}. The convective envelope was supposed to rotate as a solid body through the evolution. The transport
coefficients for chemicals associated to thermohaline and rotation-induced mixings were simply added in the diffusion
equation and the possible interactions between the two mechanisms were not considered. The rotation-induced mixing
modifies the internal chemical structure of main sequence stars, although its signatures are revealed only later in
the stellar evolution. These models lie closer to the observational data but still not close enough. 

As an alternative to the pure $^3{\rm He}$-driven thermohaline convection, the model of magneto-thermohaline 
mixing was proposed by \citet{Denissenkov2009}.  On the basis of three-dimensional numerical simulations of 
thermohaline convection, it was suggested that the salt-finger 
spectrum might be shifted towards larger diameters by a toroidal magnetic field (\citealt{Denissenkov2011}).
     
\citet{Wachlin2011} have computed full evolutionary sequences of red giant branch stars close to the luminosity bump and 
also found that thermohaline mixing is not efficient enough for fingering convection to reach the bottom of the convective envelope 
of red giants. To reach the contact, the diffusion coefficient has to be artificially increased by about four orders of 
magnitude.

\section{Summary}

With this work we have added two more open clusters with carbon-to-nitrogen and  carbon isotope ratios determined from 
high-resolution spectra to the study of evolutionary dominated changes of chemical composition in 
low mass stars.
The average value of $^{12}{\rm C}/^{13}{\rm C}$\ isotope ratios in four giants of Cr\,261 is equal to $18\pm 2$ and to $12\pm 1$ in 
   two clump stars; it is equal to $16\pm 1$ in four clump stars of the open cluster NGC\,6253. The mean C/N ratios in Cr\,261 and 
   NGC\,6253 are equal to $1.67\pm 0.06$ and  $1.37\pm 0.09$, respectively. 
   The cluster Cr\,261 provided the data point 
   for stars with the lowest turn-off mass ($1.1~M_{\odot}$), while the cluster  NGC\,6253 -- for most metal-rich stars 
   (${\rm [Fe/H]}=+0.46$). 

The $^{12}{\rm C}/^{13}{\rm C}$ and C/N values in Cr\,261 and NGC\,6253  within limits of uncertainties agree 
with the theoretical model of cool-bottom processing (\citealt{Boothroyd1999}) and with the recently computed model 
of thermohaline-induced mixing  (\citealt{Charbonnel2010}).

\begin{acknowledgements}
This research has made use of SIMBAD (operated at CDS, Strasbourg, France), VALD (\citealt{Kupka2000}) and NASA's Astrophysics Data System. 
Bertrand Plez (University of Montpellier II) and Guillermo Gonzalez  
(Washington State University) were particularly generous in providing us with 
atomic data for CN and C$_2$ molecules, respectively. We thank the referee for a careful reading of the manuscript and 
constructive suggestions for the improvement of the paper.
\end{acknowledgements}


\begin{thebibliography}{}

 \bibitem[Angelou et al.(2012)]{Angelou2012} Angelou, G.~C., Stancliffe, R.~J., Church, R.~P., Lattanzio, J.~C., \& Smith, G.~H.\ 2012, arXiv:1202.2859 
 \bibitem[Anthony-Twarog et al.(2010)]{Anthony-Twarog2010} Anthony-Twarog, B.~J., Deliyannis, C.~P., Twarog, B.~A., Cummings, J.~D., \& Maderak, R.~M.\ 2010, \aj, 139, 2034 
\bibitem[Barisevi{\v c}ius et al.(2011)]{Barisevicius2011} Barisevi{\v c}ius, G., Tautvai{\v s}ien{\.e}, G., Berdyugina, S., Chorniy, Y., \& Ilyin, I.\ 2011, Baltic Astronomy, 20, 53 
\bibitem[Barisevi{\v c}ius et al.(2010)]{Barisevicius2010} Barisevi{\v c}ius, G., Tautvai{\v s}ien{\.e}, G., Berdyugina, S., Chorniy, Y., \& Ilyin, I.\ 2010, Baltic Astronomy, 19, 157 
 \bibitem[Bensby et al.(2004)]{Bensby2004} Bensby, T., Feltzing, S., \& Lundstr{\"o}m, I.\ 2004, \aap, 415, 155 
 \bibitem[Boothroyd \& Sackmann(1999)]{Boothroyd1999} Boothroyd, A.~I., \& Sackmann, I.-J.\ 1999, \apj, 510, 232 
\bibitem[Boothroyd et al.(1995)]{Boothroyd1995} Boothroyd, A.~I., Sackmann, I.-J., \& Wasserburg, G.~J.\ 1995, \apjl, 442, L21 
 \bibitem[Bragaglia et al.(1997)]{Bragaglia1997} Bragaglia, A., Tessicini, G., Tosi, M., Marconi, G., \& Munari, U.\ 1997, \mnras, 284, 477 
 \bibitem[Bragaglia \& Tosi(2006)]{Bragaglia2006} Bragaglia, A., \& Tosi, M.\ 2006, \aj, 131, 1544 
 \bibitem[Cantiello \& Langer(2010)]{Cantiello2010} Cantiello, M., \& Langer, N.\ 2010, \aap, 521, A9 
 \bibitem[Carraro et al.(1998)]{Carraro1998} Carraro, G., Ng, Y.~K., \& Portinari, L.\ 1998, \mnras, 296, 1045 
 \bibitem[Carretta et al.(2007)]{Carretta2007} Carretta, E., Bragaglia, A., \& Gratton, R.~G.\ 2007, \aap, 473, 129 
 \bibitem[Carretta et al.(2005)]{Carretta2005} Carretta, E., Bragaglia, A., Gratton, R.~G., \& Tosi, M.\ 2005, \aap, 441, 131 
 \bibitem[Chanam{\'e} et al.(2005)]{Chaname2005} Chanam{\'e}, J., Pinsonneault, M., \& Terndrup, D.~M.\ 2005, \apj, 631, 540 
 \bibitem[Charbonnel(2006)]{Charbonnel2006} Charbonnel, C.\ 2006, EAS Publications Series, 19, 125 
 \bibitem[Charbonnel(1994)]{Charbonnel1994} Charbonnel, C.\ 1994, \aap, 282, 811 
 \bibitem[Charbonnel \& Lagarde(2010)]{Charbonnel2010} Charbonnel, C., \& Lagarde, N.\ 2010, \aap, 522, A10 
 \bibitem[Charbonnel \& Zahn(2007)]{Charbonnel2007} Charbonnel, C., \& Zahn, J.-P.\ 2007, \aap, 467, L15 
 \bibitem[Charbonnel et al.(1998)]{Charbonnel1998} Charbonnel, C., Brown, J.~A., \& Wallerstein, G.\ 1998, \aap, 332, 204 
 \bibitem[Coplen et al. (2002)]{Coplen2002} Coplen T.\ et al.\ 2002, Pure \& Appl.\ Chem.\ 74:1987-2017 
\bibitem[Denissenkov(2010)]{Denissenkov2010} Denissenkov, P.~A.\ 2010, \apj, 723, 563 
\bibitem[Denissenkov \& Merryfield(2011)]{Denissenkov2011} Denissenkov, P.~A., \& Merryfield, W.~J.\ 2011, \apjl, 727, L8 
\bibitem[Denissenkov et al.(2009)]{Denissenkov2009} Denissenkov, P.~A., Pinsonneault, M., \& MacGregor, K.~B.\ 2009, \apj, 696, 1823 
 \bibitem[De Silva et al.(2007)]{DeSilva2007} De Silva, G.~M., Freeman, K.~C., Asplund, M., et al.\ 2007, \aj, 133, 1161 
 \bibitem[Eggleton et al.(2006)]{Eggleton2006} Eggleton, P.~P., Dearborn, D.~S.~P., \& Lattanzio, J.~C.\ 2006, Science, 314, 1580 
 \bibitem[Friel et al.(2003)]{Friel2003} Friel, E.~D., Jacobson, H.~R., Barrett, E., et al.\ 2003, \aj, 126, 2372 
 \bibitem[Gaige(1993)]{Gaige1993} Gaige, Y.\ 1993, \aap, 269, 267 
 \bibitem[Gilroy(1989)]{Gilroy1989} Gilroy, K.~K.\ 1989, \apj, 347, 835 
 \bibitem[Gilroy \& Brown(1991)]{Gilroy1991} Gilroy, K.~K., \& Brown, J.~A.\ 1991, \apj, 371, 578 
 \bibitem[Gozzoli et al.(1996)]{Gozzoli1996} Gozzoli, E., Tosi, M., Marconi, G., \& Bragaglia, A.\ 1996, \mnras, 283, 66 
 \bibitem[Gratton et  al.(2006)]{gratton06} Gratton, R., Bragaglia, A., Carretta, E.,  Tosi, M., 2006, \apj, 642, 462 
 \bibitem[Gratton et al.(2000)]{Gratton2000} Gratton, R.~G., Sneden, C., Carretta, E., \& Bragaglia, A.\ 2000, \aap, 354, 169 
\bibitem[Gratton(1988)]{Gratton1988} Gratton, R.~G.\ 1988, Rome Obs. Preprint Ser., 29
 \bibitem[Grevesse \& Sauval(2000)]{Grevesse2000} Grevesse, N., \& Sauval, A.~J.\ 2000, Origin of Elements in the Solar System, Implications of Post-1957 Observations, 261 
\bibitem[Gurtovenko \& Kostyk(1989)]{Gurtovenko1989} Gurtovenko, E.~A., \& Kostyk, R.~I.\ 1989, Fraunhofer’s Spectrum and a System of Solar Oscillator Strengths, Kiev: Nauk.~dumka, 1989, p. 200
 \bibitem[Gustafsson et al.(1999)]{Gustafsson1999} Gustafsson, B., Karlsson, T., Olsson, E., Edvardsson, B., \& Ryde, N.\ 1999, \aap, 342, 426 
 \bibitem[Iben(1965)]{Iben1965} Iben, I., Jr.\ 1965, \apj, 142, 1447 
 \bibitem[Janes \& Phelps(1994)]{Janes1994} Janes, K.~A., \& Phelps, R.~L.\ 1994, \aj, 108, 1773 
 \bibitem[Johansson et al.(2003)]{Johansson2003} Johansson, S., Litz{\'e}n, U., Lundberg, H., \& Zhang, Z.\ 2003, \apjl, 584, L107 
\bibitem[Kupka et al.(2000)]{Kupka2000} Kupka, F.~G., Ryabchikova, T.~A., Piskunov, N.~E., Stempels, H.~C., \& Weiss, W.~W.\ 2000, Baltic Astronomy, 9, 590 \bibitem[Kunze(2003)]{Kunze2003} Kunze, E.\ 2003, Progress in Oceanography, 56, 399 
\bibitem[Kurucz(2005)]{Kurucz2005} Kurucz, R.~L.\ 2005, Memorie della Societa Astronomica Italiana Supplementi, 8, 189 
 \bibitem[Kurucz(1993)]{Kurucz1993} Kurucz, R.\ 1993, ATLAS9 Stellar Atmosphere Programs and 2 km/s grid.~Kurucz CD-ROM No.~13.~ Cambridge, Mass.: Smithsonian Astrophysical Observatory, 1993  
\bibitem[Lagarde et al.(2011)]{Lagarde2011} Lagarde, N., Charbonnel, C., Decressin, T., \& Hagelberg, J.\ 2011, \aap, 536, A28 
 \bibitem[Luck(1994)]{Luck1994} Luck, R.~E.\ 1994, \apjs, 91, 309 
 \bibitem[Mazur et al.(1995)]{Mazur1995MNRAS} Mazur, B., Krzeminski, W., \& Kaluzny, J.\ 1995, \mnras, 273, 59 
 \bibitem[Mikolaitis et al.(2011b)]{Mikolaitis2011B} Mikolaitis, {\v S}., Tautvai{\v s}ien{\.e}, G., Gratton, R., Bragaglia, A., \& Carretta, E.\ 2011b, \mnras, 416, 1092 
 \bibitem[Mikolaitis et al.(2011a)]{Mikolaitis2011A} Mikolaitis, {\v S}., Tautvai{\v s}ien{\.e}, G., Gratton, R., Bragaglia, A., \& Carretta, E.\ 2011a, \mnras, 413, 2199 
 \bibitem[Mikolaitis et al.(2010)]{Mikolaitis2010} Mikolaitis, {\v S}., Tautvai{\v s}ien{\.e}, G., Gratton, R., Bragaglia, A., \& Carretta, E.\ 2010, \mnras, 407, 1866 (Paper~I) 
 \bibitem[Montalto et al.(2009)]{Montalto2009} Montalto, M., Piotto, G., Desidera, S., et al.\ 2009, \aap, 505, 1129 
\bibitem[Palmerini et al.(2011)]{Palmerini2011} Palmerini, S., La Cognata, M., Cristallo, S., \& Busso, M.\ 2011, \apj, 729, 3 
 \bibitem[Pace, Pasquini, \& Fran{\c c}ois(2008)]{pace08} Pace, G., Pasquini, L., Fran{\c  c}ois, P., 2008, \aap, 489, 403 
 \bibitem[Phelps et al.(1994)]{Phelps1994} Phelps, R.~L., Janes, K.~A., \& Montgomery, K.~A.\ 1994, \aj, 107, 1079 
 \bibitem[Piatti et al.(1998)]{Piatti1998} Piatti, A.~E., Clari{\'a}, J.~J., Bica, E., Geisler, D., \& Minniti, D.\ 1998, \aj, 116, 801 
\bibitem[Radko(2010)]{Radko2010} Radko, T.\ 2010, Journal of Fluid Mechanics, 645, 121 
\bibitem[Ruddick(2003)]{Ruddick2003} Ruddick, B.\ 2003, Progress in Oceanography, 56, 483 
 \bibitem[Sestito et al.(2008)]{Sestito2008} Sestito, P., Bragaglia, A., Randich, S., et al.\ 2008, \aap, 488, 943 
 \bibitem[Sestito et al.(2007)]{Sestito2007} Sestito, P., Randich, S., \& Bragaglia, A.\ 2007, \aap, 465, 185 
 \bibitem[Shi et al.(2002)]{Shi2002} Shi, J.~R., Zhao, G., \& Chen, Y.~Q.\ 2002, \aap, 381, 982 
\bibitem[Schmitt(2003)]{Schmitt2003} Schmitt, R.\ 2003, Progress in Oceanography, 56, 419 
\bibitem[Schmitt(1983)]{Schmitt1983} Schmitt, R.~W.\ 1983, Physics of Fluids, 26, 2373 
 \bibitem[Smiljanic et al.(2009)]{Smiljanic2009} Smiljanic, R., Gauderon, R., North, P., et al.\ 2009, \aap, 502, 267 
 \bibitem[Stern(1960)]{Stern1960} Stern, M.~E.\ 1960, Tellus, 12, 172 
\bibitem[Tautvai{\v s}ien{\.e} et al.(2010a)]{Tautvaisiene2010a} Tautvai{\v s}ien{\.e}, G., Barisevi{\v c}ius, G., Berdyugina, S., Chorniy, Y., \& Ilyin, I.\ 2010, Baltic Astronomy, 19, 95 
\bibitem[Tautvai{\v s}ien{\.e} et al.(2010b)]{Tautvaisiene2010b} Tautvai{\v s}ien{\.e}, G., Edvardsson, B., Puzeras, E., Barisevi{\v c}ius, G., \& Ilyin, I.\ 2010, \mnras, 409, 1213 
 \bibitem[Tautvai{\v s}ien{\.e} et al.(2005)]{Tautvaisiene2005} Tautvai{\v s}ien{\.e}, G., Edvardsson, B., Puzeras, E., \& Ilyin, I.\ 2005, \aap, 431, 933 
\bibitem[Tautvai{\v s}ien{\.e} et al.(2001)]{Tautvaisiene2001} Tautvai{\v s}ien{\.e}, G., Edvardsson, B., Tuominen, I., \& Ilyin, I.\ 2001, \aap, 380, 578 
 \bibitem[Tautvai{\v s}iene et al.(2000)]{Tautvaisiene2000} Tautvai{\v s}iene, G., Edvardsson, B., Tuominen, I., \& Ilyin, I.\ 2000, \aap, 360, 499 
 \bibitem[Ulrich(1972)]{Ulrich1972} Ulrich, R.~K.\ 1972, \apj, 172, 165 
\bibitem[Wachlin et al.(2011)]{Wachlin2011} Wachlin, F.~C., Miller Bertolami, M.~M., \& Althaus, L.~G.\ 2011, \aap, 533, A139 
\bibitem[Wasserburg et al.(1995)]{Wasserburg1995} Wasserburg, G.~J., Boothroyd, A.~I., \& Sackmann, I.-J.\ 1995, \apjl, 447, L37 

\end{thebibliography}
\end{document}